# Electronic structure of a Si-containing topological Dirac semimetal CaAl$_2$Si$_2$

Tao Deng[1,2,3,*], Cheng Chen[3,4,5*], Hao Su[3*], Junyi He[6], Aiji Liang[3,5], Shengtao Cui[3], Haifeng Yang[3], Chengwei Wang[1,2,3], Kui Huang[3], Chris Jozwiak[4], Aaron Bostwick[4], Eli Rotenberg[4], Donghui Lu[4,7], Makoto Hashimoto[4,7], Lexian Yang[8,9], Zhi Liu[1,3], Yanfeng Guo[3], Gang Xu[6], Zhongkai Liu[3,5†], and Yulin Chen[3,5,8,10‡]

[1]*CAS Center for Excellence in Superconducting Electronics (CENSE), State Key Laboratory of Functional Materials for Informatics, Shanghai Institute of Microsystem and Information Technology (SIMIT), Chinese Academy of Sciences, Shanghai 200050, P. R. China*

[2]*University of Chinese Academy of Sciences, Beijing 100049, P. R. China*

[3]*School of Physical Science and Technology, ShanghaiTech University, Shanghai 201210, P. R. China*

[4]*Advanced Light Source, Lawrence Berkeley National Laboratory, Berkeley, CA 94720, USA*

[5]*ShanghaiTech Laboratory for Topological Physics, ShanghaiTech University, Shanghai 201210, China*

[6]*Wuhan National High Magnetic Field Center & School of Physics, Huazhong University of Science and Technology, Wuhan 430074, P. R. China*

[7]*Stanford Synchrotron Radiation Lightsource, SLAC National Accelerator Laboratory, Menlo Park, California 94025, USA*

[8]*Collaborative Innovation Center of Quantum Matter, Beijing 100084, P. R. China*

[9]*State Key Laboratory of Low Dimensional Quantum Physics, Department of Physics and Collaborative Innovation Center of Quantum Matter, Tsinghua University, Beijing 100084, P. R. China*

[10]*Department of Physics, University of Oxford, Oxford OX1 3PU, United Kingdom*

*\*These authors contributed equally to this work.*
*Email*:†*liuzhk@shanghaitech.edu.cn*,‡ *yulin.chen@physics.ox.ac.uk*

**There has been an upsurge in the discovery of topological quantum materials, where various topological insulators and semimetals have been theoretically predicted and experimentally observed. However, only very few of them contains silicon, the most widely used element in electronic industry. Recently, ternary compound CaAl$_2$Si$_2$ has been predicted to be a topological Dirac semimetal, hosting Lorentz-symmetry-violating quasiparticles with a strongly tilted conical band dispersion. In this work, by using high-resolution angle-resolved photoemission spectroscopy (ARPES), we investigated the comprehensive electronic structure of CaAl$_2$Si$_2$. A pair of topological Dirac crossings is observed along the $k_z$ direction, in good agreement with the *ab initio* calculations,**

**confirming the topological Dirac semimetal nature of the compound. Our study expands the topological material family on Si-containing compounds, which have great application potential in realizing low-cost, nontoxic electronic device with topological quantum states.**

PACS：71.55.Ak, 71.20.-b, 73.20.-r, 79.60.-i

# I. INTRODUCTION

The last several years have witnessed the intensive research on the topological quantum materials (TQMs), represented by the discoveries of topological insulators (TIs) and topological semimetals (TSMs).[1-5] Various interesting topologically non-trivial quasiparticles in the TQMs have been identified both theoretically and experimentally, including 2D Dirac fermions on the surface of TIs (e.g., $Bi_2Te_3$[5] and $Bi_2Se_3$[6]), 3D Dirac/Weyl fermions [Fig. 1(a)] in the bulk state of topological Dirac semimetals (TDSMs, e.g., $Na_3Bi$[7-10] and $Cd_3As_2$[11-13]) and topological Weyl semimetals (TWSMs, e.g., TaAs[14-18] and NbAs[19,20]). Moreover, these quasiparticles found in TQMs could go beyond the standard model, resembling the situation in particle physics. For example, the Lorentz-symmetry-violating Dirac fermion [exhibiting strongly tilted Dirac conical dispersion, see Fig. 1(a)] could be found in type-II Dirac semimetals (e.g. 1T phase $PtSe_2$[21,22] and $PdTe_2$[23-26]), showing very different physics and application potentials.

On the other hand, most of the TQMs found so far contain toxic elements (e.g. As, Te, Se…). However, from the application point of view, the current semiconductor industry would favor the TQMs with non-toxic, more common and widely used elements (e.g. Si, Ge, Al, Ca…). Recently, the search over the material database suggests a plethora of TQMs with non-trivial topological electronic structure,[27] including many Si-containing compounds such as $BaSi_2$[27] and $TiNiSi$[28]. Nevertheless, only few have been experimental confirmed, such as ZrSiS (topological Dirac nodal line semimetal)[29,30] and CoSi (chiral semimetal with multifold fermions)[31,32]. Therefore, discovery and deep investigation on Si-containing TQMs with topological Dirac/Weyl fermions are still urged.

Recently, a ternary compound $CaAl_2Si_2$, with three most common metal elements, is theoretically predicted to host multiple topological electronic states,[33] including topological Dirac nodal lines and Lorentz-symmetry-violating Dirac fermions. Specifically, with the effect of spin-orbit coupling (SOC), the Dirac nodal lines are gapped, while the Dirac fermions are protected by the crystal symmetry, making the compound a TDSM [see Fig. 1(b)].

In this work, we have conducted a systematic study on the electronic structure of $CaAl_2Si_2$ by using high-resolution synchrotron based ARPES. Both electron and hole pockets on the Fermi Surface (FS) are identified, proving the semimetallic nature of the compound, which is in nice agreement with the transport measurements[33-35]. Moreover, the photon energy dependent measurement confirms the topological band crossing along the Γ−A direction and the Lorentz-symmetry-violating Dirac fermions are clearly identified. The measured electronic structure shows great consistency with our *ab initio* calculation. Our study expands the topological semimetal family on Si-containing materials, and shed light on the potential industry application of TQMs, with stable, low-cost, non-toxic compounds.

## II. RESULTS AND DISCUSSION

### A. Topologically non-trivial electronic structure of CaAl$_2$Si$_2$

CaAl$_2$Si$_2$ crystallizes in a trigonal La$_2$O$_3$-type structure (with $P\bar{3}m1$ space group, no.164) as illustrated in Fig. 1(c),[36] isostructural to the PtSe$_2$ family. Si and Al atoms are arranged in double-corrugated hexagonal layers, with Ca atom layers intercalated forming the hexagonal frame. The crystal structure respects both the inversion and C$_3$ rotational symmetries.

The band inversion takes place within the *p*-orbitals of Si, as illustrated in Fig. 1(b). Under the crystal-field splitting (CFS), the *p*-orbitals split into $p_z$ and $p_{x,y}$ with different energies. Due to the SOC effect and as well as the C$_3$ rotational symmetry, the in-plane $p_{x,y}$ further split into singlets with irreducible representation as $\overline{DT_4 \oplus DT_5}$ and $\overline{DT_6}$. As a result of crystal field and Si bonding, bands derived from out-of-plane $p_z$ orbitals possess a relatively stronger $k_z$-dispersion compared to those derived from $p_{x,y}$ orbitals. When the out-of-plane bandwidth becomes greater than CFS and SOC, crossings between the bands occur. As CaAl$_2$Si$_2$ crystal structure respects both the inversion and time-reversal symmetries, the first intersection of the bands ($\overline{DT_4 \oplus DT_5}$ and $\overline{DT_6}$) is fourfold degenerated. The crossing point is protected against the SOC hybridization by the C$_3$ rotational symmetry, forming a bulk topological Dirac point along the Γ−A direction [as illustrated in the 3D Brillouin Zone (BZ) in Fig. 1(d)]. The absence of magnetism in the compound could well preserve the time reversal symmetry and hence protect this TDSM state. On the other hand, the second intersection of the bands (two $\overline{DT_6}$ bands with opposite parities) is gapped by the SOC, forming a tiny band-inversion gap where topological surface states could be expected [resembling the case of Fe(Te,Se)[37,38] and LiFeAs[39]].

## B. Sample synthesis and characterization

High-quality CaAl$_2$Si$_2$ single crystals were synthesized by using a high temperature self-flux method (see Appendix for details). The crystals could be cleaved between Si-Al planes along the [001] direction. After cleavage, shiny (001) surface with sizes around several millimeters was exposed [Fig. 2(b)(i)], suitable for the photoemission study. The quality of the sample was verified by both the X-ray photoemission spectroscopy (XPS) and X-ray diffraction (XRD). The XPS spectrum shows sharp characteristic peaks of Ca *3s*, Ca *3p*, Al *2s*, Al *2p*, and Si *2p* orbitals [Fig. 2(a)] and the XRD gives clear diffraction patterns of (100), (010) and (001) surfaces [Fig. 2(b)(ii)(iii)(iv)]. The lattice constants could be deduced as a=b=4.13 Å, c=7.145 Å, consistent with the previous reports.[27,33] Fig. 2(c) illustrates a broad constant energy contour (CEC) at Fermi level ($E_F$) from our ARPES measurement, and the sixfold rotational symmetry of the spectrum confirms the cleavage surface to be (001) [the momentum directions of $k_x$, $k_y$, $k_z$ are defined to be parallel to Γ−K, Γ−M, and Γ−A respectively; see Fig. 1(d)].

## C. General electronic structure of CaAl$_2$Si$_2$

We first focus on the overall electronic structure of the CaAl$_2$Si$_2$ (Fig. 3). As is shown in Fig. 2(c), the FS consists of hole pockets at the Γ point, and small electron pockets at the M point, which could be verified by the constant energy contours (CECs) at different binding energies shown in Fig. 3(a)(i-v). The feature near Γ extends with the binding energy, while the feature near M disappears at high binding energy, in nice agreement with the theoretical calculations [Fig. 3(a)(vi-x)]. Therefore, the CaAl$_2$Si$_2$ is confirmed to be a semimetal, consistent with the previous transport measurement.[35] Moreover, complex textures appear and evolve at higher binding energies, indicating the existence of multiple electronic bands. From the high symmetry band dispersion measured with linear horizontal (LH) polarized light along M−Γ−M direction [Fig. 3(b)(i)], several hole-like bands (labeled as α, β, γ, and ξ) and one electron-like band (labeled as η) could be identified. In comparison, with linear vertical (LV) polarized light [Fig. 3(b)(ii)], β, γ, η bands seem to be suppressed, while α and another feature (labeled as ε) are enhanced due to the photoemission matrix element effect. The hole-like α band and electron-like η band cross the E$_F$, contributing to the semimetallic nature of the compound. The related calculation is plotted in Fig. 3 (c), consistent with experiment results.

### D. $k_z$ evolution and surface states in CaAl$_2$Si$_2$

To trace the evolution of the bulk electronic states along the $k_z$ direction, systematic photon energy dependent measurement was performed (energy ranging from 90 eV to 238 eV). The FS in the $k_z$-$k_y$ plane is plotted in Fig. 4(a), where clear periodic patterns was observed with a periodicity of 2π/c (BZ is indicated by the dashed lines with high symmetry points labeled). The electronic structures around the Γ and A points are illustrated by the 3D volume plots [Fig. 4(b)(i-ii)] as well as the high symmetry cuts along the M−Γ−M and L−A−L directions [Fig. 4(c-d)]. Clearly, we found that the top of α, β, γ, and ξ bands all shift downward at A point (the related calculations are overlapped on the spectrum), showing strong $k_z$ evolution which proves their bulk nature. Besides, several additional features (labeled as δ$_1$, δ$_2$, and δ$_3$) were identified that do not evolve with $k_z$. Fig. 4(e) plots the band dispersion along the $k_z$ direction at $k_y$=0.10 Å$^{-1}$ (Γ'−A') and $k_y$=0.51 Å$^{-1}$ (Γ''−A''), from which we could find the $\delta_1$ and $\delta_3$ features form dispersionless straight lines, indicating their surface nature. The corresponding slab calculation along $\overline{M}-\overline{\Gamma}-\overline{M}$ further proved the existence of these surface states [Fig. 4(f)].

The sizes of both electron and hole pockets on the FS are estimated. The electron pocket (η band) near M and the hole-pocket (α+β band) near Γ each has an area of 0.01 Å$^{-2}$ and 0.094 Å$^{-2}$, respectively. The effective mass of electron is evaluated to be m*=0.397m$_e$, closing to the value obtained in the transport measurement 0.32m$_e$ (m$_e$ denotes the free electron mass)[33].

### E. Observation of the Lorentz-symmetry-violating Dirac fermion

From the photon energy dependent data, we could examine the dispersion along the Γ−A−Γ direction. Figure 5(b)(i-ii) shows the dispersion measured with two different ranges of photon energies but both covering a full A−Γ period. Similar patterns of evolution were identified in both spectra (labeled by purple and green dashed curves): three features disperse with $k_z$, while another feature does not vary along $k_z$. By comparing with the band assignment in Fig. 4, we attribute the dispersive bands to α+β (degenerated along Γ−A−Γ), γ, ε+ξ (nearly degenerated along Γ−A−Γ), and the dispersionless band to $\delta_1$, respectively [see Supplementary Figs. S1 and S2]. We noticed that the γ and ε+ξ bands form a crossing in the middle of Γ and A (k$_z$≈0.53 π/c, marked as 'D') [see Supplementary Fig. S3]. The band crossing appears more apparent in the 2D curvature plot in Fig. 5(b)(iii), which shows nice agreement with the theoretical prediction and the schematic drawn in Fig. 1(b) and Fig. 5(a), where the tilted Dirac cone consists of γ (mostly Si $p_z$-orbital) and ε (mostly Si $p_x$+$p_y$-orbital) and the 'D' point is the predicted Dirac point. We further confirm the shape of Dirac fermion by showing the dispersion along the M'−D−M' direction (parallel to M−Γ−M, but crossing the D point) at h$v$=48eV [Fig. 5(c)(i)], where a straight Dirac conical shape is observed, in agreement with the calculation in Fig. 5(c)(ii).

### III. CONCLUSION

In conclusion, our ARPES measurement as well as the *ab initio* calculation revealed the complete electronic structure of $CaAl_2Si_2$. The nice agreement between the experiment result and the calculation strongly supports that $CaAl_2Si_2$ is a $C_3$-symmetry-protected topological Dirac semimetal. We note that the Dirac points in $CaAl_2Si_2$ are buried much deeper below $E_F$ ($E_D \approx E_F - 1.6$ eV) so that they might have little contribution to the transport properties. Our ARPES result not only confirms the topological nature of $CaAl_2Si_2$, but widens the investigation in exploring Si-containing TQM family, which has application potential in the future electronic and spintronic devices compatible with current semiconductor industry.

### ACKNOWLEDGEMENTS

We thank Dr. Cheng Chen and Dr. Aiji Liang for their assistance during the beamtime at Stanford Synchrotron Radiation Lightsource (SSRL) BL 5-2 and Advanced Light Source (ALS) BL 7.0.2; Dr. Zhengtai Liu, Dr. Dawei Shen and Dr. Wanling Liu for their assistance during the beamtime at Shanghai Synchrotron Radiation Facility (SSRF) BL03U. We acknowledge the support from SiP·ME$^2$ project under contract No. 11227902 from National Natural Science Foundation of China. The work is supported by the National Key R&D program of China (Grant No.2017YFA0305400, 2018YFA0307000 and 2017YFA0304600), the National Natural Science Foundation of China (Grant No. 11774190, No. 11874022, No. 11674229) and the Strategic Priority Research Program of Chinese Academy of Sciences (Grant No. XDA18010000). All authors contributed to the scientific planning and discussions. The authors declare no competing financial interests.

T. Deng, C. Chen and H. Su contributed equally to this work.

## APPENDIX: MATERIALS AND METHODS

### 1. Sample synthesis

The $CaAl_2Si_2$ single crystals were grown using a high temperature self-flux method[40]. Starting materials including calcium block (99.9%), aluminum sheet (99.99%) and silicon powder (99.99%) were mixed in a molar ratio of Ca : Al : Si = 1: 15: 2. The mixture was sealed in a quartz tube, heated up to 1150 °C in a furnace and kept at the temperature for 30 hours, then cooled down slowly to 1000 °C at rate of 2 K/h. The assembly was finally taken out of the furnace at 1000 °C and was put into a centrifuge immediately to remove the excess flux. Compositions of the crystals were examined by using energy-dispersive X-ray (EDX) spectroscopy. The phase and quality examinations of $CaAl_2Si_2$ were performed on a single-crystal x-ray diffractometer equipped with a Mo K$\alpha$ radioactive source ($\lambda$ = 0.71073 Å).

### 2. Angle-resolved photoemission spectroscopy

ARPES measurements mentioned above were performed at BL 7.0.2 at Adevanced Light

Source (ALS), USA, beamline 5-2 at Stanford Synchrotron Radiation Lightsource (SSRL), USA and BL03U at Shanghai Synchrotron Radiation Facility (SSRF), China. The $CaAl_2Si_2$ single crystals were first elaborately selected and well prepared in air. Then they were cleaved *in situ* along the (001) surface without spoiling, and measured in ultrahigh vaccum with a base pressure of better than $1.5\times10^{-10}$ Torr as well as a 80 K sample temperature. Data were recorded by a VG Scienta DA30 analyzer in SSRF and ALS or a D80 analyzer in SSRL. The angle resolution was 0.2° and the overall energy resolutions were better than 15 meV.

### 3. *Ab initio* calculations

The first-principles calculations based on the density function theory (DFT) were carried out by the Vienna ab initio simulation package (VASP)[41-43], in which the exchange-correlation potential was treated within the generalized gradient approximation in the form of Perdew-Burke-Ernzerhof[44]. The cutoff energy of the plane-wave basis was set as 400eV, and the Brillouin zone (BZ) was sampled by the $19\times19\times9$ meshes in the self-consistent calculations. Maximally localized Wannier functions for 3s, 3p orbitals of Al, and 3s, 3p orbitals of Si were generated by the WANNIER90 package[45]. Then the topological properties were calculated by using Wanniertools[46].

### Supplementary materials

Supplementary materials to this article can be found online or from the authors.

# Figures and legends

**Fig. 1**

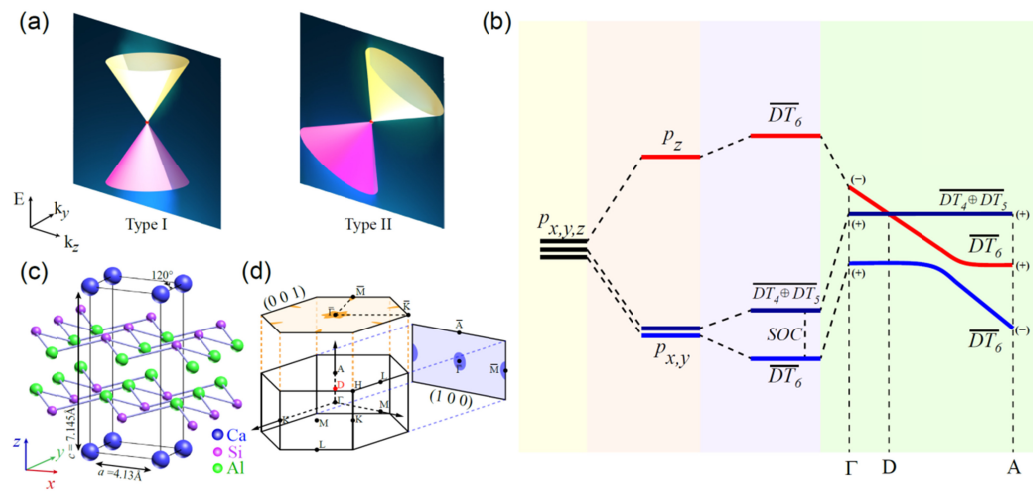

**Fig. 2**

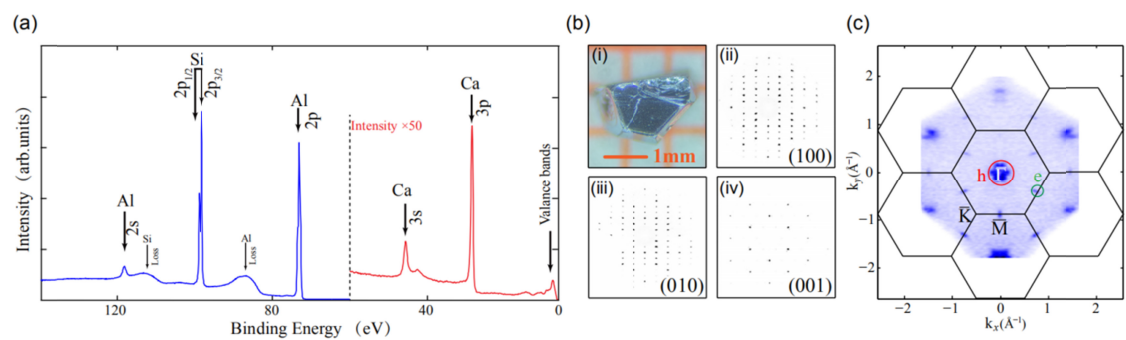

**Fig. 3**

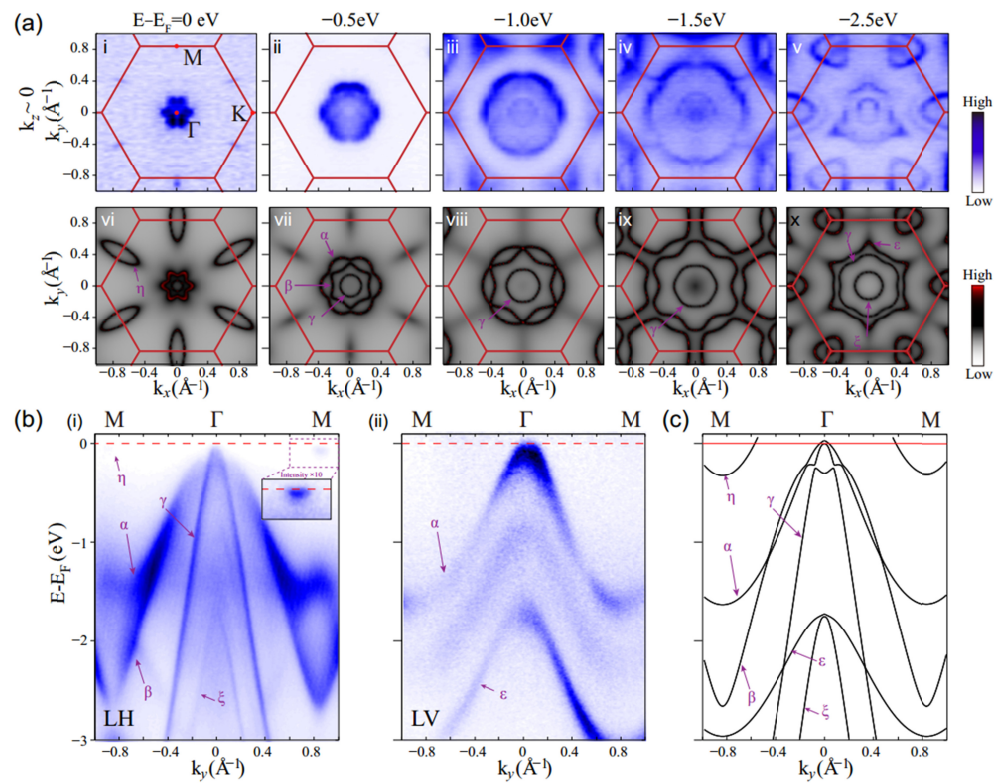

**Fig. 4**

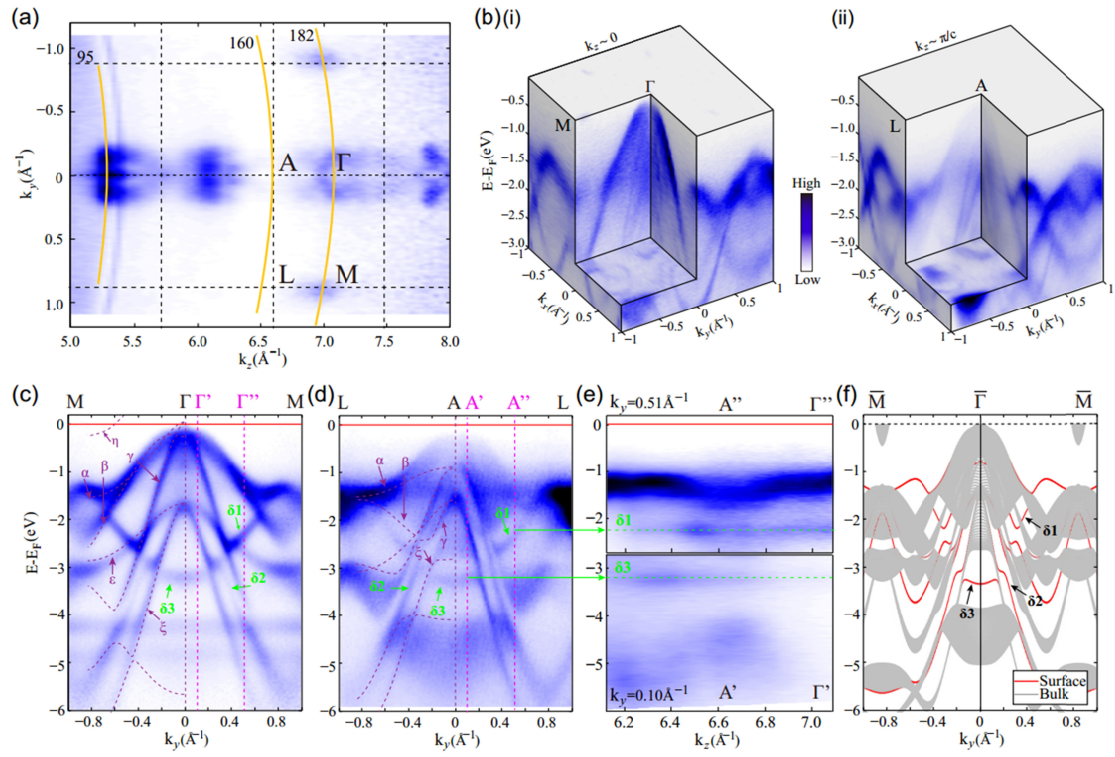

**Fig. 5**

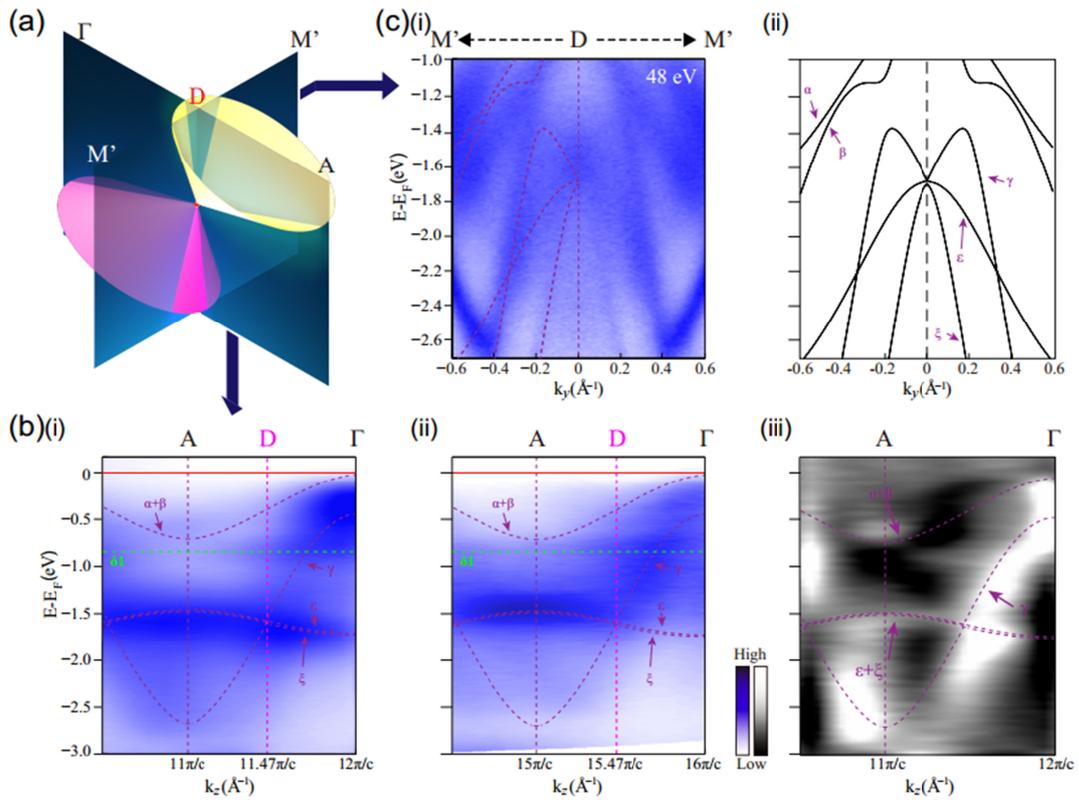

FIG. 1. Formation of the type-II Dirac Fermion in CaAl$_2$Si$_2$. (a) Schematic illustration of type-I and type-II Dirac fermions in the energy-momentum (E-**k**) space. (b) Schematic illustration of the *p*-orbital energy level evolution of *p*-orbitals forming the Dirac point. The different color represent the crystal field splitting (pink) and spin orbit coupling (purple and green). The symmetry of the energy levels are labeled with irreducible representations ($\overline{DT_4} \oplus \overline{DT_5}$ and $\overline{DT_6}$) and parity(+/−). (c) Schematic of the crystal structure of CaAl$_2$Si$_2$, with lattice constants (a, c) marked. (d) The 3D Brillouin zone (BZ) of CaAl$_2$Si$_2$ and its projected surface BZ to (001) plane (creamy) and (100) surface (blue). High-symmetry points (e.g. Γ, A, K, M etc.) and Dirac point (D) are labeled.

FIG. 2. Characterization of CaAl$_2$Si$_2$ samples. (a) The integrated core-level photoemission spectrum showing characteristic Ca, Si and Al core-level peaks. (b) (i) Photograph of the high quality CaAl$_2$Si$_2$ single crystal with flat surface used for ARPES measurements and (ii-iv) X-ray diffraction patterns from the (100), (010), (001) surface of CaAl$_2$Si$_2$. (c) A large momentum scale of photoemission Fermi energy contour taken by 182 eV linear horizontal polarized photons on the (001) surface, showing the sixfold rotational symmetry and characteristics of semimetallic FSs. Electron and hole pockets are marked by green and red circles.

FIG. 3. Overall Electronic Structure of CaAl$_2$Si$_2$. (a)(i-v) Photoemission intensity map of Constant Energy Contours (CECs) around Γ point at 0, 0.5, 1.0, 1.5, 2.5 eV below E$_F$, respectively. (vi-x) Corresponding theoretical *ab-initio* calculation showing CECs at the same energies as (i-v), respectively. Bulk bands α, β, γ, ε, ξ, η are labeled. (b) Band dispersions along high symmetry M−Γ−M directions measured with (i) linear horizontal (LH) and (ii) linear vertical (LV) polarized photons. Identified bulk bands are labeled. (c) Calculated band dispersons along high symmetry M−Γ−M directions. CECs in (a) are symmetrized according to the crystal symmetry. The integrated energy window is ±50 meV. The data were collected using photons with *hν*=182 eV (corresponding to $k_z$~0) at 80 K.

FIG. 4. $k_z$ evolution of the electronic band structure of CaAl$_2$Si$_2$. (a) CECs at the E$_F$ in the $k_z$-$k_y$ plane. The dashed rectangles indicate the (100) surface BZs lattice. The cut directions shown in (c) and (d) were indicated by the yellow curve with the photon energy used labeled. (b) 3D plots of the in-plane electronic structure at (i) $k_z$~0 (ii) $k_z$~π/c measured at 182eV and 160eV with LH polarization, respectively. (c)(d) Band dispersions along high symmetry M−Γ−M (at h$v$=95 eV, LH+LV polarization) and L−A−L (at h$v$=160 eV, LH polarization) directions. The corresponding calculation results were overlapped with dotted curves. Different bands are labeled. Γ', Γ", A', A" label the $k_y$ positions used in (e). (e) Band dispersions along A'−Γ' ($k_y$=0.10Å$^{-1}$) and A"−Γ" ($k_y$=0.51Å$^{-1}$) direction, as is labeled in (c)(d), respectively. Position of the dispersionless surface states are marked by green dashed lines. (f) Corresponding slab calculation along $\overline{M}-\overline{\Gamma}-\overline{M}$. Identified surface states ($\delta_1$, $\delta_2$ and $\delta_3$) are labeled.

FIG. 5. Observation of the Dirac fermion in CaAl$_2$Si$_2$. (a) Schematic of the Dirac fermion as well as the cross-sections along different direction. Different positions in the momentum space are labeled. (b)(i)(ii)Band dispersions along high-symmetry Γ−A−Γ direction (across different $k_z$ range) with (iii) side-by-side comparison to 2D curvature analysis results. Corresponding calculation results are overlapped with dotted curves. Different bands are labeled. The $k_z$ position of the Dirac point (D) is labeled. (c)(i) Band dispersion along M'−D−M' direction (labeled in (a)) as well as (ii) the corresponding calculated bulk bands. The data were extracted from photon energy dependent measurements from 76~95 eV (LH+LV polarization) in (b)(i) and 148~182 eV (LH polarization) in (b)(ii) at 80 K. The data were collected using photons with h$v$=48 eV (corresponding to $k_z$≈0.53 π/c) with LH+LV polarization at 13 K in (c)(i).